\documentclass{article}

\usepackage{template/waspaa23}
\usepackage{amsfonts}
\usepackage{amsmath}
\usepackage{graphics}
\usepackage{multirow}
\usepackage{hyperref}
\usepackage{subcaption}
\usepackage{pgfplots}
\pgfplotsset{compat=1.18}
\usepackage{tikz}
\usetikzlibrary{positioning}
\usepackage{ctable}
\usepackage{xcolor}
\usepackage{cite}
\usepackage{placeins}
\usepackage{ifthen}
\renewenvironment{thebibliography}[1]{
  \begin{oldthebibliography}{#1}
    \setlength{\itemsep}{0pt} 
    \setlength{\parskip}{0em}
}
{
  \end{oldthebibliography}
}

\def\ninept{\def\baselinestretch{.922}\let\normalsize\small\normalsize}
\title{Flexible multichannel speech enhancement for noise-robust frontend}
\name{Ante Juki\'{c}, Jagadeesh Balam, Boris Ginsburg}
\address{NVIDIA, USA}

\begin{document}
\ninept

\maketitle

\begin{abstract}
This paper proposes a flexible multichannel speech enhancement system with the main goal of improving robustness of automatic speech recognition (ASR) in noisy conditions.
The proposed system combines a flexible neural mask estimator applicable to different channel counts and configurations and a multichannel filter with automatic reference selection.
A transform-attend-concatenate layer is proposed to handle cross-channel information in the mask estimator, which is shown to be effective for arbitrary microphone configurations.
The presented evaluation demonstrates the effectiveness of the flexible system for several seen and unseen compact array geometries, matching the performance of fixed configuration-specific systems.
Furthermore, a significantly improved ASR performance is observed for configurations with randomly-placed microphones.
\end{abstract}

\section{Introduction}
Capturing a speech signal by distant microphones often results in a speech signal corrupted by noise and reverberation.
In addition to the desired speech signal, the recordings typically contain undesired signal components which may negatively impact intelligibility of the speech signal for a human listener~\cite{beutelmann2006prediction} or cause regression in the performance of automatic speech recognition (ASR) and related downstream tasks~\cite{yoshioka2012making}.

When multiple microphones are available, the undesired signal components can be greatly reduced using multichannel signal processing~\cite{benesty2008microphone}.
Such multichannel systems can effectively exploit the spatial diversity present in the input signals and enhance the desired signal while suppressing the noise interference, e.g, by enhancing the signal from a certain direction.
As demonstrated in several studies, multichannel filtering can effectively reduce noise and increase robustness of ASR systems~\cite{barker2017third}.
Effective processing of arbitrary multichannel signals is often critical in practical applications, e.g., if providing an ASR service without controlling the microphone configuration used to capture the speech~\cite{nvidia_riva}.

Recently, mask-based multichannel filters have been shown to be very effective for multichannel speech enhancement~\cite{souden2013multichannel, heymann2015blstm, higuchi2016robust, erdogan2016improved, zhang2022end, ochiai2023moving,zmolikova2023neural}.
A mask estimator is typically designed to predict a probability that the desired speech signal is active in a particular time-frequency (TF) bin.
This TF mask can then be used to estimate statistics necessary for the design of a multichannel filter.
The mask can be estimated by exploiting spatial, temporal or spectral information in the input signals.
For example, a mask can be estimated by clustering the directional statistics~\cite{ito2016complex, higuchi2016robust}.
However, a drawback of such approaches is that they rely dominantly on spatial information, and neural mask estimators can effectively exploit temporal and spectral information more effectively~\cite{heymann2015blstm, erdogan2016improved, haeb2020far}.
Neural mask estimators often rely only on spectro-temporal information in the input signal and are applied on a single channel~\cite{haeb2020far}.
While such an approach may be practical, exploiting the available spatial information is typically beneficial.

Generalization of neural mask estimators to multichannel scenarios has been an active area of research~\cite{yoshioka2018multi, luo_2020_tac, yemini2020scene, wang2020neural, chang2020end, zhang2022end, taherian2022one, yoshioka_2022_vararray}.
A split-apply-combine approach has been widely used, where the same model is applied to individual channels and the obtained masks are aggregated~\cite{erdogan2016improved,chang2020end, zhang2022end}.
However, this approach does not take into account the cross-channel information and applying the same model repeatedly may be inefficient.
A transform-average-concatenate (TAC) layer has been proposed in~\cite{luo_2020_tac} for handling arbitrary permutation and count of channels.
In~\cite{yoshioka_2022_vararray}, TAC has been combined with Conformer~\cite{gulati_2020_conformer} blocks in a continuous speech separation system.
An inter-channel attention layer has been used in~\cite{wang2020neural} to deal with arbitrary channel configurations for two-source separation.
However, only magnitude features have been considered and performance for different array configurations and noise robustness has not been evaluated.
Personalized multichannel speech enhancement for arbitrary compact arrays has been considered in~\cite{taherian2022one}.
However, the enhancement has been performed by masking the reference microphone instead of applying multichannel filtering.

The mask estimator considered here consists of an alternating sequence of channel and temporal blocks.
The temporal blocks are based on the Conformer architecture~\cite{gulati_2020_conformer}, similar to recent studies studies~\cite{koizumi_2021_dfconf, yoshioka_2022_vararray}.
The contribution of this work is threefold.
Firstly, we propose to use a transform-attend-concatenate layer, which is a generalization of TAC~\cite{luo_2020_tac} and the inter-channel attention layer from~\cite{wang2020neural}.
As opposed to fixed averaging in TAC, using attention is more effective at handling arbitrary microphone configurations.
Similarly, we use an attention-based channel reduction layer instead of fixed average pooling.
Secondly, as opposed to~\cite{taherian2022one}, a minimum-variance distortionless response (MVDR) beamformer with automatic reference channel selection is used at train time.
Thirdly, we evaluate the proposed system on several seen and unseen compact microphone array configurations and on microphone configurations with randomly-placed microphones.
The presented results demonstrate the effectiveness of the considered system in various scenarios.
\section{Signal model}
\label{sec:signal_model}
Assuming a single static speaker is recorded in a room with an array of $M$ microphones, the time-domain signal $y_m(t)$ at the $m$-th microphone at time $t$ can be written as
\begin{equation}
\label{eq:signal_model}
  y_m(t) = h_m(t) \ast s(t) + v_m(t) ,
\end{equation}
where $h_m(t)$ is the room impulse response (RIR), $s(t)$ is the clean speech, $v_m(t)$ is the additive noise, and $\ast$ denotes convolution.
The signal model in~\eqref{eq:signal_model} can be approximated in the short-time Fourier transform (STFT) domain as
\begin{equation}
  y_m(f,n) = h_m(f,n) \ast s(f,n) + v_m(f,n) ,
\end{equation}
where $f$ is the frequency subband index and $n$ is the time frame index~\cite{avargel2007system}.
The goal of this work is to obtain a time-domain estimate of the early part of the speech signal image $d(f,n) = h_{r,\mathrm{early}}(f,n) \ast s(f,n)$ at the reference microphone $r$ from the multichannel signal $\mathbf{y}(f,n) = \left[ y_1(f,n), \dots, y_M(f,n) \right]^\mathrm{T}$.
Note that the reference microphone $r$ is not assumed to be set or known in advance and we assume batch processing.
\section{Model architecture}
\label{sec:model_architecture}
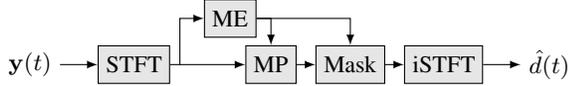
\begin{figure}[h]
  \centering
  \begin{tikzpicture}
    \node[draw,
      rectangle,
      minimum height=0.5cm,
      fill=black!10
    ] (stft) at (0,0){STFT};
    \node[draw,
      rectangle,
      minimum height=0.5cm,
      fill=black!10,
      above right=0.10cm and 0.45cm of stft,
      align=center
    ] (mask_estimator) {ME};
    \node[draw,
      rectangle,
      minimum height=0.5cm,
      fill=black!10,
      right=1.0cm of stft,
      align=center
    ] (mc_processor) {MP};
    \node[draw,
      rectangle,
      minimum height=0.5cm,
      fill=black!10,
      right=0.25cm of mc_processor,
      align=center
    ] (mask) {Mask};
    \node[draw,
      rectangle,
      minimum height=0.5cm,
      fill=black!10,
      right=0.25cm of mask,
      align=center
    ] (istft) {iSTFT};
    \node[left=0.5cm of stft] (input) {$\mathbf{y}(t)$};
    \node[right=0.5cm of istft] (output) {$\hat{d}(t)$};
    \node[right=0.5pt of stft] (stft_connect) {};
    \draw[-latex] (input) -- (stft);
    \draw[-latex] (stft) -- (mc_processor);
    \draw[-latex] (stft_connect.center) |- (mask_estimator);
    \draw[-latex] (mask_estimator) -| (mc_processor);
    \draw[-latex] (mask_estimator) -| (mask);
    \draw[-latex] (mc_processor) -- (mask);
    \draw[-latex] (mask) -- (istft);
    \draw[-latex] (istft) -- (output);
  \end{tikzpicture}
  \caption{Processing system including mask estimation (ME), multichannel processing (MP) and single-channel masking (Mask).}
  \label{fig:block_scheme}
\end{figure}

We consider a typical mask-driven multichannel processing system as depicted in the block diagram in Fig.~\ref{fig:block_scheme}~\cite{boeddeker2018front}.
A real-valued mask is estimated from the multichannel input using a neural model and used in multichannel processing to estimate the desired speech signal and select the reference microphone, and additional masking may be applied on the single-channel output.
\subsection{Mask estimator}
\label{subsec:mask_estimator}
\begin{figure}
  \centering
  \scriptsize
  \begin{tikzpicture}
    \node[draw,
      rectangle,
      minimum width=2.5cm,
      minimum height=0.4cm,
      fill=black!10,
      align=center
    ] (channel_block_1) {channel block};
    \node[draw,
      rectangle,
      minimum width=2.5cm,
      minimum height=0.4cm,
      fill=black!10,
      below=0.4cm of channel_block_1,
      align=center
    ] (temporal_block_1) {temporal block};
    \draw[-latex] (channel_block_1) -- (temporal_block_1) node[pos=0.5,right]{$M$-ch};
    \node[below=0.1cm of temporal_block_1] (before_channel_reduction) {};
    \draw[-] (temporal_block_1) -- (before_channel_reduction.center);
    \node[draw,
      rectangle,
      minimum width=2.5cm,
      minimum height=0.4cm,
      fill=black!10,
      below=0.1cm of before_channel_reduction,
      align=center
    ] (channel_reduction) {channel reduction};
    \draw[-latex] (before_channel_reduction.center) -- (channel_reduction);
    \node[draw,
      rectangle,
      minimum width=2.5cm,
      minimum height=0.4cm,
      fill=black!10,
      below=0.4cm of channel_reduction,
      align=center
    ] (temporal_block_2) {temporal block};
    \draw[-latex] (channel_reduction) -- (temporal_block_2) node[pos=0.5,right]{1-ch};
    \node[below=0.1cm of temporal_block_2] (before_nonlinearity) {};
    \draw[-] (temporal_block_2) -- (before_nonlinearity.center);
    \node[draw,
      rectangle,
      minimum width=2.5cm,
      minimum height=0.4cm,
      fill=black!10,
      below=0.1cm of before_nonlinearity,
      align=center
    ] (nonlinearity) {$\sigma$};
    \draw[-latex] (before_nonlinearity.center) -- (nonlinearity) ;
    \node[right=0.75cm of nonlinearity] (output_mask) {$\mathbf{g}(n) \in \mathbb{R}^F$};
    \draw[-latex] (nonlinearity) -- (output_mask) ;
    \node[left=0.75cm of channel_block_1] (input) {$\mathbf{Z}(n) \in \mathbb{R}^{2F \times M}$} ;
    \draw[-latex] (input) -- (channel_block_1) ;
    \node[draw=gray,
      dotted,
      rectangle,
      minimum height=1.5cm,
      minimum width=3.5cm,
      below=0.2cm of channel_block_1,
      anchor=center
    ] (repeat_multi_channel) {};
    \node[fill=white] at (repeat_multi_channel.east) (repeat_multi_channel_label) {\scriptsize repeat} ;
    \node[draw=gray,
      dotted,
      rectangle,
      minimum height=0.62cm,
      minimum width=3.5cm,
      anchor=center
    ] at (temporal_block_2) (repeat_single_channel) {};
    \node[fill=white] at (repeat_single_channel.east) (repeat_single_channel_label) {\scriptsize repeat} ;
  \end{tikzpicture}
  \caption{Block scheme of the mask estimator. Input features are defined in~\eqref{eq:features}, $\sigma$ is a sigmoid unit, and the output is a TF mask.}
  \label{fig:block_mask_estimator}
\end{figure}
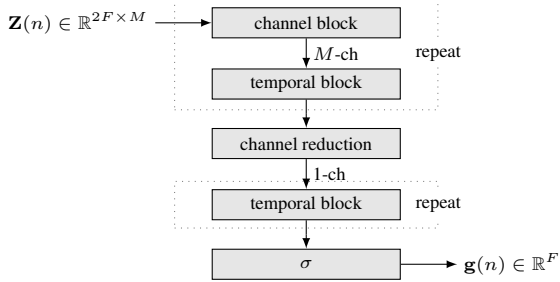

The first step in building an effective multichannel mask estimator is to select relevant features~\cite{yoshioka2018multi}.
It has been demonstrated in~\cite{yoshioka2018multi,yoshioka2018recognizing} that inter-channel phase differences (IPDs) are quite effective when combined with magnitude features.
Therefore, similarly as in ~\cite{yoshioka2018multi, yoshioka_2022_vararray}, we use a combination of magnitude and IPD features as an input to the neural mask estimator.
More specifically, we use $\mathbf{Z}(n) = \left[ \mathbf{Z}_\mathrm{mag}^\mathrm{T}(n), \mathbf{Z}_\mathrm{IPD}^\mathrm{T}(n) \right]^\mathrm{T} \in \mathbb{R}^{2F \times M}$ as input features, with
\begin{equation}
\label{eq:features}
\resizebox{0.91\hsize}{!}{$%
  \left\lbrace \mathbf{Z}_\mathrm{mag}(n) \right\rbrace_{f,m} = \left| y_m(f,n) \right| ,
  \,
  \left\lbrace \mathbf{Z}_\mathrm{IPD}(n) \right\rbrace_{f,m} = \angle \frac{y_m(f,n)}{\bar{y}(f,n)} ,
$}
\end{equation}
where $\bar{y}(f,n) = \sum_m y_m(f,n) / M$ is the complex-valued channel-averaged spectrum. Similarly to~\cite{yoshioka2018multi}, magnitude features were mean and variance normalized, while only mean normalization was applied to IPD features.

The mask estimator depicted in Fig.~\ref{fig:block_mask_estimator} consists of an alternating sequence of blocks operating over time and channel, similarly as in~\cite{taherian2022one, yoshioka_2022_vararray}.
Temporal blocks are using Conformer~\cite{gulati_2020_conformer} to transform an $N$-frame input sequence into an $N$-frame output sequence, similarly to~\cite{yoshioka_2022_vararray}.
After the last temporal layer, a sigmoid is used to obtain the TF mask $\mathbf{g}(n)$.
A channel block transforms $M$ input streams into $M$ output streams and needs to support any number of input channels.
In~\cite{yoshioka_2022_vararray}, a transform-average-concatenate (TAC) layer~\cite{luo_2020_tac} was used to combine information across the channels by averaging.
Here, we propose to use a transform-attend-concatenate channel block, i.e., to replace the fixed averaging with cross-channel attention.
Assuming $\mathbf{Z}(n) \in \mathbb{R}^{K \times M}$ is an $M$-channel input of the proposed channel block at time step $n$, the output $\tilde{\mathbf{Z}}(n) \in \mathbb{R}^{\tilde{K} \times M}$ can be written
\begin{equation}
  \label{eq:transform_attend_concatenate}
  \tilde{\mathbf{Z}}(n) =
    \begin{bmatrix}
      \mathrm{ReLU} \left( \mathbf{W}_\mathrm{C} \mathbf{Z}(n) \right) \\
      \mathrm{MHSA} \left( \mathrm{ReLU} \left( \mathbf{W}_\mathrm{A} \mathbf{Z}(n) \right) \right)
    \end{bmatrix}
\end{equation}
where $\mathbf{W}_\mathrm{C}, \mathbf{W}_\mathrm{A}$ are linear transforms and $\mathrm{MHSA} : \mathbb{R}^{K \times M} \to \mathbb{R}^{\tilde{K} \times M}$ is a multi-head self-attention block~\cite{vaswani2017attention}.
Similarly, we propose to replace fixed channel-wise averaging~\cite{yoshioka_2022_vararray} with a squeeze-and-excite-like attention weighting~\cite{hu2018squeeze}.
As earlier, assuming $\mathbf{Z}(n)$ is an input at time step $n$, the output $\check{\mathbf{z}}(n) \in \mathbb{R}^{K}$ can be written as
\begin{equation}
  \check{\mathbf{z}}(n) = \mathbf{Z}(n)~\mathrm{softmax} \left( \mathbf{V}^\mathrm{T} \mathbf{Q} \mathbf{1} M^{-1} \right),
\end{equation}
where $\mathbf{Q} = \mathbf{W}_Q \bar{\mathbf{Z}},~ \mathbf{V} = \mathbf{W}_V \bar{\mathbf{Z}} \in \mathbb{R}^{K \times M}$, $\bar{\mathbf{Z}} = \sum_n \mathbf{Z}(n) / N$ is the temporal average of the input, and multiplication with $\mathbf{1}M^{-1}$ from right denotes averaging across the $M$ columns.
\subsection{Multichannel processing}
The obtained TF mask $g(f,n)$ is used to estimate the covariance matrix in subband $f$ for the desired speech $\mathbf{\Phi}_{dd}(f) = \sum_n g(f,n) \mathbf{y}(f,n) \mathbf{y}^\mathrm{H}(f,n) / \sum_n g(f,n)$ and the undesired signal $\mathbf{\Phi}_{uu}(f) = \sum_n g_u(f,n) \mathbf{y}(f,n) \mathbf{y}^\mathrm{H}(f,n) / \sum_n g_u(f,n)$ where $g_u(f,n) = 1 - g(f,n)$.
We use a standard MVDR formulation~\cite{souden2009optimal}
\begin{equation}
  \label{eq:mvdr}
  \mathbf{w}_{r}(f) = \frac{\mathbf{\Phi}_{uu}^{-1}(f)\mathbf{\Phi}_{dd}(f)}{\mathrm{trace} \left\lbrace \mathbf{\Phi}_{uu}^{-1}(f)\mathbf{\Phi}_{dd}(f) \right\rbrace} \mathbf{e}_{r} ,
\end{equation}
where $\mathbf{e}_{r} \in \mathbb{R}^M$ is the $r$-th column of an $M \times M$ identity matrix.
The reference channel $r$ is estimated by maximizing the average output SNR~\cite{erdogan2016improved, boeddeker2018front} as
\begin{equation}
  \label{eq:ref_estimator}
  \hat{r} = \arg\max_m \frac{\sum_f \mathbf{w}_m^H(f) \mathbf{\Phi}_{dd}(f) \mathbf{w_m}(f)}{\sum_f \mathbf{w}_m^H(f) \mathbf{\Phi}_{uu}(f) \mathbf{w_m}(f)}
\end{equation}
Optionally, a straight-through estimator can be used to propagate the gradient through~\eqref{eq:ref_estimator}~\cite{bengio2013estimating}.
The output of the multichannel filter is finally obtained as $\hat{d}(f,n) = \mathbf{w}_{\hat{r}}^\mathrm{H}(f) \mathbf{y}(f, n)$.
Additional TF masking is applied on $\hat{d}(f,n)$ by multiplying with $\max \left( g(f,n), g_\mathrm{min} \right)$, where $g_\mathrm{min}$ is a minimum mask threshold.
\subsection{Loss function}
As shown in~\cite{boeddeker_2021_cisdr}, the negative convolution-invariant signal-to-distortion ratio (SDR) is an appropriate loss function for multichannel speech enhancement.
Let $\mathbf{s} \in \mathbb{R}^T$ denote the clean speech signal reference and $\hat{\mathbf{d}} \in \mathbb{R}^T$ the estimated desired speech.
The loss function can then be written as~\cite{boeddeker_2021_cisdr}:
\begin{equation}
  \label{eq:loss}
  \mathcal{L}\left( \hat{\mathbf{d}}, \mathbf{s} \right) = -10 \log_{10} \frac{\| \hat{\mathbf{h}} \ast \mathbf{s} \|_2^2}{\| \hat{\mathbf{h}} \ast \mathbf{s} - \hat{\mathbf{d}} \|_2^2 + \alpha \|\hat{\mathbf{h}} \ast \mathbf{s}\|_2^2} ,
\end{equation}
where $\hat{\mathbf{h}} \in \mathbb{R}^L$ is the filter obtained by minimizing $\| \mathbf{h} \ast \mathbf{s} - \hat{\mathbf{d}} \|_2$ and $\alpha = 10^{-\mathrm{SDR}_\mathrm{max}/10}$ is a soft threshold~\cite{wisdom2020unsupervised}.
\section{Experimental results}
\label{sec:experiments}
\subsection{Datasets}
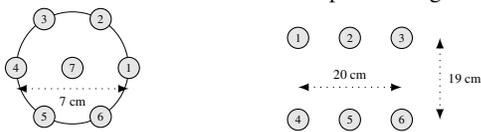
\begin{figure}[b]
  \centering
  \begin{tikzpicture}
    \node[draw,
      very thin,
      circle,
      fill=none,
      scale=5
    ] (circle) at (0,0) {};
    \node[draw,
      circle,
      fill=black!10!white,
      scale=0.5
    ] (mic_7) at (0,0) {7};
    \node[draw,
      circle,
      fill=black!10!white,
      scale=0.5
    ] (mic_1) at (0.75,0) {1};
    \node[draw,
      circle,
      fill=black!10!white,
      scale=0.5
    ] (mic_2) at (0.375,0.649) {2};
    \node[draw,
      circle,
      fill=black!10!white,
      scale=0.5
    ] (mic_3) at (-0.375,0.649) {3};
    \node[draw,
      circle,
      fill=black!10!white,
      scale=0.5
    ] (mic_4) at (-0.75,0) {4};
    \node[draw,
      circle,
      fill=black!10!white,
      scale=0.5
    ] (mic_5) at (-0.375,-0.649) {5};
    \node[draw,
      circle,
      fill=black!10!white,
      scale=0.5
    ] (mic_6) at (0.375,-0.649) {6};
    \node[below=0.02cm of mic_4] (mic_x_1) {};
    \node[below=0.02cm of mic_1] (mic_x_2) {};
    \draw[latex-latex,thin,dotted] (mic_x_1.center) -- (mic_x_2.center) node[pos=0.5,below]{\tiny 7~cm};
    \node[draw,
      circle,
      fill=black!10!white,
      scale=0.5
    ] (rec_1) at (3,0.4) {1};   
    \node[draw,
      circle,
      fill=black!10!white,
      scale=0.5,
      right=0.4cm of rec_1
    ] (rec_2) {2};
    \node[draw,
      circle,
      fill=black!10!white,
      scale=0.5,
      right=0.4cm of rec_2
    ] (rec_3) {3};
    \node[draw,
      circle,
      fill=black!10!white,
      scale=0.5,
      below=0.8cm of rec_1
    ] (rec_4) {4};
    \node[draw,
      circle,
      fill=black!10!white,
      scale=0.5,
      below=0.8cm of rec_2
    ] (rec_5) {5};
    \node[draw,
      circle,
      fill=black!10!white,
      scale=0.5,
      below=0.8cm of rec_3
    ] (rec_6) {6};
    \node[right=0.25cm of rec_3] (rec_y_1) {};
    \node[right=0.25cm of rec_6] (rec_y_2) {};
    \draw[latex-latex,thin,dotted] (rec_y_1.center) -- (rec_y_2.center) node[pos=0.5,right]{\tiny 19~cm};
    \node[below=0.4cm of rec_1] (rec_x_1) {};
    \node[below=0.4cm of rec_3] (rec_x_2) {};
    \draw[latex-latex,thin,dotted] (rec_x_1.center) -- (rec_x_2.center) node[pos=0.5,above]{\tiny 20~cm};
  \end{tikzpicture}
  \caption{Circular and rectangular microphone array configurations.}
  \label{fig:mic_arrays}
\end{figure}
Room impulse responses for several microphone configurations were generated using the image method~\cite{scheibler_2018_pyroom} in 10000, 200, 200 different rooms for train, dev and test sets with 20 source positions in each room.
Room width, length and height were randomized in ranges~$[3, 7]$\,m, $[3,9]$\,m and $[2.3, 3.5]$\,m.
A compact 7-channel circular microphone configuration with 7\,cm diameter was simulated, with six microphones evenly spaced on the circle and a single microphone in the center.
A compact 6-channel rectangular microphone configuration was simulated similar to~\cite{barker2017third}.
The circular and the rectangular configuration are depicted in Fig.~\ref{fig:mic_arrays}.
Random microphone configurations were also simulated, with six microphones placed at arbitrary locations inside each room.
For the two compact configurations, the array was placed in the horizontal plane and the array center was constrained to height in $[1,1.5]$m, while individual microphones were constrained to the same height range for the random configuration.
In all cases, microphones were placed at least 0.5\,m from the closest wall.
A sound source was simulated at a randomized height in $[1.4, 1.8]$\,m and at least 0.5\,m from the closest wall.
The reverberation time was randomized in the range $[0.1, 0.5]$\,s.
The microphone signals were simulated using clean speech from LibriSpeech~\cite{panayotov2015librispeech} and noise from \mbox{CHiME-3}~\cite{barker2017third} and DNS challenge~\cite{dubey2022icassp}.
Half of the examples in the simulated datasets included only diffuse noise, simulated as spherically-isotropic noise using~\cite{habets2008generating} with reverberant signal-to-noise ratio (RSNR) uniformly selected in $[-5, 20]$dB.
The other half of the examples included diffuse noise and one to three directional noise sources, with the RSNR for the directional sources uniformly selected in $[-5,20]$dB.
Approximately 200\,h of audio for the training set and 10\,h for dev and test sets was generated for each microphone configuration.
The sampling rate in all experiments was 16\,kHz.
\subsection{Experimental setup}
The STFT is computed with 32\,ms frame size and 16\,ms overlap.
The mask estimator includes six temporal blocks using Conformer encoder with four attention heads, hidden size 128 and convolution size 31.
The first five blocks each have five layers, while the last temporal block before the sigmoid has a single layer and $F$ output features.
The channel reduction block was placed after the third temporal block.
The baseline systems without cross-channel attention (denoted as "Att. --") use TAC channel block and average pooling for channel reduction as in~\cite{yoshioka_2022_vararray}.
The proposed systems with cross-channel attention (denoted as "Att. $\checkmark$") use attention-based layers from Section~\ref{subsec:mask_estimator}.
Linear transforms in~\eqref{eq:transform_attend_concatenate} are halving the number of features, so the total number of features after concatenation remains the same as at the input.
All configurations had approximately 10.7\,M parameters.
The matrix $\mathbf{\Phi}_{uu}(f)$ in~\eqref{eq:mvdr} is regularized by adding its trace scaled with $10^{-6}$ to its diagonal.
The loss in~\eqref{eq:loss} is calculated using the clean signal at the closest microphone as $\mathbf{s}$, a 32\,ms filter $\mathbf{h}$ and $\mathrm{SDR}_\mathrm{max} = -30$\,dB~\cite{wisdom2020unsupervised}.
The model is implemented in NVIDIA's NeMo toolkit~\cite{nvidia_nemo_toolkit}.
Training is performed on four-second audio segments with the global batch size of 192.
The model is trained for 150 epochs using adamw optimizer with learning rate $10^{-3}$ and cosine annealing scheduler with $10^4$ warmup steps and all evaluated models are obtained by averaging each parameter over the five checkpoints with the lowest dev set loss.

Flexible models are trained on either the circular dataset or on a combination of circular and random datasets.
The number of channels for each mini batch was selected randomly between 2 and 6, aiming to provide a variety of configurations in training.
The performance is evaluated on two seen compact arrays (based on the circular configuration), three unseen compact arrays (based on the rectangular configuration) and two random configurations.
For each compact configuration, we train a fixed model which is a baseline model without the proposed cross-channel attention and trained with the exact channel count and layout as used in the respective test set.

The performance is evaluated in terms of word error rate (WER), SDR\cite{vincent_2006_sdr} and short-time objective intelligibility (STOI)\cite{taal_2011_stoi}.
For both signal-based metrics, the reference signal is the clean speech signal at the closest microphone.
For WER evaluation, we used NVIDIA's \mbox{Conformer-CTC-Large} English ASR model.
Note that the the \mbox{pre-trained} ASR model has not been finetuned on the noisy or processed data used in these experiments.
\begin{table}
\caption{Performance for seen compact array configurations. "Att." denotes attention in channel blocks, "Ref. grad" denotes gradient propagation in ~\eqref{eq:ref_estimator} and $g_\mathrm{min}$ denotes the single-channel mask threshold (0\,dB means the mask is always 1, i.e., it is not applied).}
\label{table:seen_array}
\vspace{-0.2cm}
\begin{subtable}{\columnwidth}
\begin{center}
  \scriptsize
  \setlength\tabcolsep{3pt} 
  \begin{tabular}{ccccrrrr}
    \toprule
    System      & Train set & Att.       & Ref. grad  & $g_\mathrm{min}$ & SDR/dB$\uparrow$ & STOI$\uparrow$ & WER/\%$\downarrow$ \\
    \toprule
    closest mic &           &            &            &         &  1.06 & 0.65 & 32.22 \\ \midrule
    fixed 3-ch  & matched   & --         & --         &  0\,dB  &  7.71 & 0.70 & 20.74 \\
                &           & --         & --         &  -6\,dB &  9.84 & 0.72 & 20.61 \\ \midrule
    flexible    & circular  & --         & --         &  0\,dB  &  7.69 & 0.71 & 20.58 \\
                &           & --         & --         &  -6\,dB &  9.74 & 0.73 & 19.70 \\ \midrule
    flexible    & circular  & \checkmark & --         &  0\,dB  &  7.66 & 0.71 & 20.31 \\
                &           & \checkmark & --         &  -6\,dB &  9.73 & 0.73 & 19.84 \\ \midrule
    flexible    & circ+rand & --         & --         &  0\,dB  &  7.65 & 0.71 & 20.59 \\
                &           & --         & --         &  -6\,dB &  9.81 & 0.72 & 20.68 \\ \midrule
    flexible    & circ+rand & \checkmark & --         &  0\,dB  &  7.26 & 0.71 & 20.25 \\
                &           & \checkmark & --         &  -6\,dB &  9.59 & 0.73 & 20.08 \\ \midrule
    flexible    & circ+rand & \checkmark & \checkmark &  0\,dB  &  7.04 & 0.71 & 19.88 \\
                &           & \checkmark & \checkmark &  -6\,dB &  9.51 & 0.73 & 19.97 \\
    \bottomrule
  \end{tabular}
  \caption{3-ch: channels \{1,7,4\} from the circular array in Fig.~\ref{fig:mic_arrays}.}
  \label{tab:seen_3ch}
\end{center}
\end{subtable}
\smallskip
\begin{subtable}{\columnwidth}
\begin{center}
  \scriptsize
  \setlength\tabcolsep{3pt} 
  \begin{tabular}{ccccrrrr}
    \toprule
    System      & Train set & Att.       & Ref. grad  & $g_\mathrm{min}$ & SDR/dB~$\uparrow$ & STOI~$\uparrow$ & WER/\%~$\downarrow$ \\
    \toprule
    closest mic &           &            &            &         &  1.11 & 0.65 & 32.21 \\ \midrule
    fixed 6-ch  & matched   & --         & --         &  0\,dB  & 10.31 & 0.75 & 13.88 \\
                &           & --         & --         &  -6\,dB & 11.98 & 0.77 & 14.22 \\ \midrule
    flexible    & circular  & --         & --         &  0\,dB  & 10.30 & 0.75 & 13.89 \\
                &           & --         & --         &  -6\,dB & 11.90 & 0.76 & 14.14 \\ \midrule
    flexible    & circular  & \checkmark & --         &  0\,dB  & 10.25 & 0.75 & 13.98 \\
                &           & \checkmark & --         &  -6\,dB & 11.88 & 0.76 & 14.46 \\ \midrule
    flexible    & circ+rand & --         & --         &  0\,dB  & 10.18 & 0.75 & 14.05 \\
                &           & --         & --         &  -6\,dB & 11.93 & 0.76 & 15.19 \\ \midrule
    flexible    & circ+rand & \checkmark & --         &  0\,dB  &  9.73 & 0.75 & 13.78 \\
                &           & \checkmark & --         &  -6\,dB & 11.68 & 0.77 & 14.06 \\ \midrule
    flexible    & circ+rand & \checkmark & \checkmark &  0\,dB  &  9.53 & 0.75 & 13.70 \\
                &           & \checkmark & \checkmark &  -6\,dB & 11.61 & 0.77 & 14.47 \\
  \bottomrule
  \end{tabular}
  \caption{6-ch: channels 1--6 from the circular array in Fig.~\ref{fig:mic_arrays}.}
  \label{tab:seen_6ch}
\end{center}
\end{subtable}
\end{table}
\begin{table*}
\caption{Performance for unseen compact array configurations using subset of the rectangular array in Fig.~\ref{fig:mic_arrays} with channels \{1,4\} for the 2-ch configuration, channels \{1,2,4,5\} for the 4-ch configuration, channels \{1,\dots,6\} for the 6-ch configuration.}
\label{table:unseen_array}
\vspace{-0.6cm}
\begin{center}
  \scriptsize
  \begin{tabular}{cccccrrrrrrrrrr}
    \toprule
                &           &            &                     & \multicolumn{3}{c}{2-ch} & \multicolumn{3}{c}{4-ch} & \multicolumn{3}{c}{6-ch} \\ \cmidrule(lr){5-7}\cmidrule(lr){8-10}\cmidrule(lr){11-13}
    System      & Train set & Att.       & Ref. grad           & SDR/dB~$\uparrow$ & STOI~$\uparrow$ & WER/\%~$\downarrow$ & SDR/dB~$\uparrow$ & STOI~$\uparrow$ & WER/\%~$\downarrow$& SDR/dB~$\uparrow$ & STOI~$\uparrow$ & WER/\%~$\downarrow$ \\
    \midrule
    closest mic &           &            &                      &  0.96 & 0.64 & 31.64 &  1.08 & 0.65 & 30.91 &  1.21 & 0.65 & 30.46 \\ \midrule
    fixed       & matched   & --         & --                   &  6.04 & 0.68 & 24.78 &  9.13 & 0.71 & 19.31 & 10.69 & 0.73 & 15.35 \\ \midrule 
    flexible    & circular  & --         & --                   &  5.46 & 0.68 & 24.11 &  8.62 & 0.72 & 17.33 & 10.14 & 0.72 & 14.30 \\ \midrule 
    flexible    & circular  & \checkmark & --                   &  5.36 & 0.68 & 24.19 &  8.48 & 0.72 & 17.61 &  9.93 & 0.73 & 14.62 \\ \midrule 
    flexible    & circ+rand & --         & --                   &  5.94 & 0.67 & 25.36 &  8.88 & 0.71 & 18.31 & 10.42 & 0.73 & 15.19 \\ \midrule 
    flexible    & circ+rand & \checkmark & --                   &  5.46 & 0.67 & 24.87 &  8.50 & 0.72 & 17.89 & 10.02 & 0.74 & 14.50 \\ \midrule 
    flexible    & circ+rand & \checkmark & \checkmark           &  5.25 & 0.68 & 24.43 &  8.36 & 0.72 & 17.34 &  9.90 & 0.74 & 13.94 \\
    \bottomrule
  \end{tabular}
\end{center}  
\end{table*}

\begin{table}
\caption{Performance for unseen microphone configurations with randomly-placed microphones.}
\vspace{-0.6cm}
\label{table:random_array}
\begin{center}
  \scriptsize
  \setlength\tabcolsep{2.5pt} 
  \begin{tabular}{cccccrrrrrr}\toprule
                &           &            &            & \multicolumn{3}{c}{3-ch} & \multicolumn{3}{c}{6-ch} \\ \cmidrule(lr){5-7}\cmidrule(lr){8-10}
    System      & Train set & Att.       & Ref. grad            & SDR & STOI & WER & SDR & STOI & WER \\ \midrule
    closest mic &           &            &                      &  1.53 & 0.67 & 28.58 &  2.07 & 0.69 & 26.89 \\ \midrule
    flexible    & circular  & --         & --                   &  4.50 & 0.68 & 23.96 &  5.85 & 0.68 & 22.35 \\ \midrule 
    flexible    & circular  & \checkmark & --                   &  3.75 & 0.66 & 25.58 &  4.59 & 0.66 & 24.98 \\ \midrule 
    flexible    & circ+rand & --         & --                   &  7.23 & 0.67 & 24.52 &  9.38 & 0.69 & 20.89 \\ \midrule 
    flexible    & circ+rand & \checkmark & --                   &  6.78 & 0.69 & 23.33 &  8.92 & 0.72 & 18.20 \\ \midrule 
    flexible    & circ+rand & \checkmark & \checkmark           &  6.64 & 0.69 & 22.55 &  8.79 & 0.72 & 17.80 \\
    \bottomrule
  \end{tabular}
\end{center}
\end{table}

\begin{figure}[hbt]
\pgfplotstableread[col sep=comma]{tables/snr_breakdown_rectangular_6ch.csv}\snrbreakdown
\begin{tikzpicture}
  \begin{axis}[
      width=8.5cm,
      height=3.82cm,
      title=Rectangular 6-ch configuration,
      title style={yshift=-1.5ex},
      xlabel style={yshift=1.5ex},
      ylabel={WER~/~\%},
      ybar,
      bar width=5pt,
      ymajorgrids=true,
      yminorgrids=true,
      minor y tick num=3,
      minor grid style={dotted},
      xtick pos=bottom,
      legend pos=north east,
      legend style={legend columns=3, nodes={scale=0.75}},
      legend image code/.code={\draw [#1] (0cm,-0.1cm) rectangle (0.2cm,0.1cm); },
      title style={font=\scriptsize},
      label style={font=\scriptsize},
      tick label style={font=\scriptsize},
      xtick align=inside,
      xmin=-5,
      xmax=20,
      enlarge x limits = {abs=2.0}
    ]
    \addplot[draw=none,fill=black!100!white] table[x=RSNR, y=mic]{\snrbreakdown};
    \addplot[draw=none,fill=black!60!white] table[x=RSNR, y=flex_circ]{\snrbreakdown};
    \addplot[draw=none,fill=black!30!white] table[x=RSNR, y=flex_circ_rand_ref_grad]{\snrbreakdown};
    \legend{mic,flexible-circ,flexible-circ+rand+att+ref-grad}
  \end{axis}
\end{tikzpicture}
%
\pgfplotstableread[col sep=comma]{tables/snr_breakdown_random_6ch.csv}\snrbreakdown
\begin{tikzpicture}
  \begin{axis}[
      width=8.5cm,
      height=3.82cm,
      title=Random 6-ch configuration,
      title style={yshift=-1.5ex},
      xlabel={RSNR~/~dB},
      xlabel style={yshift=1.3ex},
      ylabel={WER~/~\%},
      ybar,
      bar width=5pt,
      ymajorgrids=true,
      yminorgrids=true,
      minor y tick num=3,
      minor grid style={dotted},
      xtick pos=bottom,
      legend pos=north east,
      legend style={legend columns=3, nodes={scale=0.75}},
      legend image code/.code={\draw [#1] (0cm,-0.1cm) rectangle (0.2cm,0.1cm); },
      title style={font=\scriptsize},
      label style={font=\scriptsize},
      tick label style={font=\scriptsize},
      xtick align=inside,
      xmin=-5,
      xmax=20,
      enlarge x limits = {abs=2.0}
    ]
    \addplot[draw=none,fill=black!100!white] table[x=RSNR, y=mic]{\snrbreakdown};
    \addplot[draw=none,fill=black!60!white] table[x=RSNR, y=flex_circ]{\snrbreakdown};
    \addplot[draw=none,fill=black!30!white] table[x=RSNR, y=flex_circ_rand_ref_grad]{\snrbreakdown};
    \legend{mic,flexible-circ,flexible-circ+rand+att+ref-grad}
  \end{axis}
\end{tikzpicture}
\vspace{-0.3cm}
\caption{WER in diffuse noise for rectangular and random 6-ch microphone configurations. Masking was not used.}
\label{fig:rsnr_breakdown}
\end{figure}
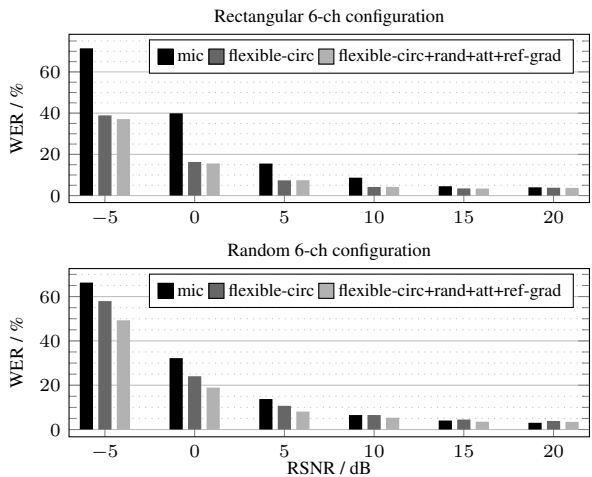

\subsection{Results}
Table~\ref{table:seen_array} shows the performance of the considered systems on test sets with microphone configurations seen by the flexible models.
The results show that all systems are able to significantly improve metrics when compared with the signal captured by the closest microphone.
The considered flexible systems achieve a performance comparable to the matched fixed system in both signal-based metrics and WER.
It can be observed that adding single-channel masking with $g_\mathrm{min}=-6$\,dB improves SDR but causes regression in WER for the 6-ch configuration due to additional speech distortions~\cite{iwamoto22_interspeech}.
Decreasing $g_\mathrm{min}$ to lower values (e.g., -20\,dB) further improved SDR but caused regression in WER, so we exclude masking from the remaining results.
Using the proposed cross-channel attention does not have a large impact on flexible models trained on the circular configuration, which is expected due to the compactness of the array used for training.
Furthermore, flexible system trained on a combination of circular and random configurations achieves the best WER when using the cross-channel attention and reference gradient propagation, with the final result on par or slightly better than the matched fixed configuration.

Table~\ref{table:unseen_array} shows the performance of the considered systems on test sets with microphone configurations unseen by the flexible models.
Across the three unseen configurations, the flexible systems achieve a comparable performance in STOI and 1\,dB or less regression in SDR compared to the matched baselines.
In terms of WER, the best result is achieved by flexible systems.
The results obtained on the 2-ch configuration is comparable to the matched fixed configuration.
The results on the flexible 4-ch configurations are better by up to almost 2\% absolute WER.
The best result on the 6-ch configuration is achieved by the flexible system trained on a combination of circular and random configurations with the proposed cross-channel attention and reference gradient propagation.

Table~\ref{table:random_array} shows the performance of the considered systems on tests set with unseen microphone configurations with randomly-placed microphones.
It can be observed that the flexible system trained on the circular configuration improves performance in all metrics compared to the closest microphone.
Training a flexible system on a combination of circular and random configurations performs better, especially in terms of SDR, as the model is exposed to more unseen microphone configurations.
Furthermore, using the proposed cross-channel attention and reference gradient propagation improves the performance in WER, achieving improvements of more than 1.5\% absolute WER for the 3-ch configuration and more than 3\% absolute WER for the 6-ch configuration compared to the baselines without the cross-channel attention.

Finally, a breakdown of the performance in terms of WER for a scenario with diffuse noise is shown in Fig.~\ref{fig:rsnr_breakdown}.
In can be observed that the proposed system with cross-channel attention and reference gradient propagation is consistently better than the baseline system across the considered conditions.
\section{Conclusions}
\label{sec:conclusions}
In this paper, we introduced a flexible multichannel speech enhancement system suitable for improving robustness of ASR systems in noisy conditions with arbitrary microphone configurations.
The presented model includes a mask estimator with attention-based transform-attend-concatenate channel blocks and channel reduction, and automatic reference channel selection.
The system was evaluated in several scenarios with seen and unseen microphone configurations to compare its performance against array-specific models.
A good speech enhancement performance was observed on both seen and unseen compact arrays, with the results comparable to array-specific models.
Furthermore, the proposed system outperforms the baseline system for scenarios with unconstrained configurations with random microphone placement.
Several challenges remain open for future investigations, such as online processing and applications to enhancement and separation with ad-hoc microphone configurations.
\FloatBarrier
\cleardoublepage
\bibliographystyle{template/IEEEtran}
\bibliography{main}
\end{document}